\documentclass[pra,twocolumn,groupedaddress,showpacs,floatfix]{revtex4}

\usepackage{graphics}
\usepackage{graphicx}
\usepackage{bm}
\usepackage{amsmath}
\usepackage{amsfonts}
\usepackage{amssymb}
\usepackage{latexsym}

\begin{document}

\title{Energy concentration in composite quantum systems}
\author{Andreas Kurcz,$^1$ Antonio Capolupo,$^2$ Almut Beige,$^1$ Emilio Del
  Giudice,$^3$ and Giuseppe Vitiello$\, ^2$}
\affiliation{$^1$The School of Physics and Astronomy, University of Leeds, Leeds, LS2 9JT, United Kingdom}
\affiliation{$^2$Dipartimento di Matematica e Informatica and \\
I.N.F.N., Universit\'a di Salerno,~Fisciano~(SA)-84084,~Italy}
\affiliation{$^3$I.N.F.N. - via Celoria 16 - Milano, Italy}

\date{\today}

\begin{abstract}
The spontaneous emission of photons from optical cavities and from trapped atoms has been studied extensively in the framework of quantum optics. Theoretical predictions based on the rotating wave approximation (RWA) are in general in very good agreement with experimental findings. However, current experiments aim at combining better and better cavities with large numbers of tightly confined atoms. Here we predict an energy concentrating mechanism in the behavior of such a composite quantum system which cannot be described by the RWA. Its result is the continuous leakage of photons through the cavity mirrors, even in the absence of external driving. We conclude with a discussion of the predicted phenomenon in the context of thermodynamics.
\end{abstract}

\maketitle

\section{Introduction}

In quantum systems with purely unitary evolutions, the total energy is a conserved quantity, since the corresponding Hamiltonian always commutes with itself. However, this argument does not apply to  systems with Hamiltonians which are self-adjoint but not Hermitian \cite{Kurcz_NJP} and to open quantum systems \cite{Bartana}. The spontaneous emission of a photon is always related to the loss of energy from its source \cite{Hegerfeldt93,Dalibard,Carmichael}. For example, in laser sideband cooling, a red-detuned laser field excites an electronic state of a strongly confined ion via the annihilation of a phonon from its motion. When followed by spontaneous photon emission, the phonon is permanently lost \cite{Wineland}. Contrarily to common believe, we show here that even un-excited and un-driven quantum systems might constantly leak energy into their environment. The origin of the predicted effect are non-zero decay rates and the counter-rotating terms in the interaction Hamiltonian which are usually neglected as part of the RWA.

Although the validity of this approximation has been questioned in the past \cite{agarwal,knight}, it is commonly used to describe quantum optical systems. An exception is Hegerfeldt \cite{hegerfeldt2}, who shows that the counter-rotating terms in the interaction between two atoms and the free radiation field can result in a small violation of Einstein's causality. Zheng {\em et al.}~\cite{Zubairy} also avoided the RWA and predicted corrections to the spontaneous decay rate of a single atom at very short times. Recently, Werlang {\em et al.}~\cite{Werlang} pointed out that it might be possible to obtain photons by simply placing an atom inside an optical cavity but no quantitative predictions have been made and no justification for the relevant master equation has been given. When avoiding the RWA, it has to be avoided everywhere, also in the system-bath interaction.

This paper contains a rigorous derivation of the quantum optical master equation for bosonic systems which uses only the Born and the dipole approximation. We apply our results to individual quantum systems (trapped atoms and optical cavities), and to a composite quantum system consisting of many atoms inside an optical resonator. It is shown that for sufficiently large numbers of atoms inside the cavity, the stationary state photon emission rate can be as large as typical detector dark count rates. Its parameter dependence could be verified experimentally using currently available atom-cavity systems \cite{Trupke,Reichel,Esslinger}. Our calculations confirm the relevance of the effects predicted in \cite{Werlang} when the atom-cavity interactions are collectively enhanced \cite{Holstein,Shah}. Similar energy concentrating effects might contribute significantly to the sudden heating in sonoluminescence experiments \cite{SL} and are responsible for temperature limits in cooling experiments \cite{review,cool}.

There are five sections in this paper. Section \ref{Model} contains a rigorous derivation of the master equation of a single bosonic quantum system beyond the validity range of the RWA. Section \ref{single} calculates the corresponding stationary photon emission rate. Assuming that the photon emission in the absence of external driving remains negligible in this case, we find that two of the constants in this master equation are approximately zero. Section \ref{comp} uses an analogously derived master equation to calculate the stationary state cavity photon emission rate for a composite quantum system consisting of many atoms inside an optical resonator. For feasible experimental parameters, we predict stationary state emission rates as high as $300$ photons per second. A detailed discussion of our results can be found in Section \ref{conc}. 

\section{Master equation of a single quantum system beyond the RWA} \label{Model}

Let us begin by studying individual quantum systems, like optical cavities and tightly confined atoms, with the ability to emit photons. Their Hamiltonian $H$ in the Schr\"odinger picture and in the Born and the dipole approximation can be written as $H = H_0 + H_{\rm int}$ with \footnote{This form of $H_{\rm int}$ avoids fixing phase factors of states.}
\begin{eqnarray} \label{H}
H_0 &=& \hbar \omega ~s^+ s^- + \sum _{{\bf k},\lambda} \hbar \omega _k ~ a ^\dagger _{{\bf k} \lambda} a _{{\bf k} \lambda}~, \nonumber \\
H_{\rm int} &=& \sum_{{\bf k}, \lambda} \hbar ~ \big( g_{{\bf k} \lambda} ~ a _{{\bf k} \lambda} + \tilde g_{{\bf k} \lambda} ~ a _{{\bf k} \lambda}^\dagger \big) ~ s^+ + {\rm h.c.} 
\end{eqnarray}
and $|\tilde g_{{\bf k} \lambda} | = |g_{{\bf k} \lambda}|$. The free radiation field consists of an infinite number of one-dimensional harmonic oscillators with wave vectors {\bf k}, frequencies $\omega_k$, polarisations $\lambda$, annihilation operators $a_{{\bf k} \lambda}$, and coupling constants $g _{{\bf k} \lambda}$ and $\tilde g _{{\bf k} \lambda}$. In case of an optical cavity, $\omega \equiv \omega_{\rm c}$ is the frequency of its field mode and the $s^\pm$ are the photon creation and annihilation boson operators $c^\dagger$ and $c$:
\begin{equation} \label{bose}
\left [ c, c^\dagger \right] =1 ~, ~~ \left [ c^\dagger, c^\dagger \right] = 0 = \left [ c, c \right] ~,
\end{equation}
In case of a large number of tightly confined two-level atoms with states $|0 \rangle $ and $| 1 \rangle $, $\hbar \omega \equiv \hbar \omega_0$ is the energy of the excited state $|1 \rangle$ of a single atom and the $s^\pm$ are the collective raising and lowering operators $S^\pm$ with
\begin{equation} \label{Dicke4}
\left [ S^-, S^+ \right ] = 1 ~, ~~ \left [ S^+, S^+ \right] = 0 = \left [ S^-, S^- \right] ~,
\end{equation}
as we shall see in the next paragraph.

Suppose the atoms are confined in a region with linear dimensions that are much smaller than the wavelength of the emitted light. Then $|{\bf k} \cdot ( {\bf r}_j - {\bf r}_i )| \ll 1$ for most particle positions ${\bf r}_i$ and ${\bf r}_j$ and for a wide range of wave vectors ${\bf k}$. This implies that all particles experience approximately the same $g _{{\bf k} \lambda}$ and $\tilde g _{{\bf k} \lambda}$. The Hamiltonian $H$ remains therefore the same, if we replace $s^+s^-$ in $H_0$ by $\sigma_3$ and $s^\pm$ in $H_{\rm int}$ by $\sigma^\pm$. Here $\sigma^\pm$ and $\sigma_3$ are defined as 
\begin{eqnarray} \label{sigma}
\sigma^{\pm} \equiv \sum _{i=1} ^N \sigma _i ^{\pm} ~, ~~
\sigma _3 \equiv \sum_{i=1} ^N \sigma _{3i} 
\end{eqnarray}
with $\sigma _{3i} = {1 \over 2} \left ( | 1 \rangle_{ii} \langle 1 |-| 0 \rangle_{ii} \langle 0 | \right )$, $\sigma_i^+ = | 1 \rangle_{ii} \langle 0 |$ and $\sigma_i ^- = | 0 \rangle_{ii} \langle 1 |$ being the su(2) spin-like operators of atom $i$:
\begin{equation} \label{su2}
\left [ \sigma _{3i} , \sigma_i ^{\pm} \right ] = \pm \sigma_i ^ {\pm} \, , ~ \left [ \sigma_i ^-  , \sigma_i ^+\right] = - 2 \sigma _{3i} ~ .
\end{equation}
If the atoms are initially all in their ground state, they evolve under the
action of the operators (\ref{sigma}) into the Dicke-symmetric states:
\begin{eqnarray}
| l \rangle _{\rm p} &\equiv& \left [ | 0_1 0_2 0_3 \dots 0_{N-l} 1_{N-l+1} 1_{N-l+2} \dots 1 \rangle + \dots \right . \nonumber\\
 && \left . + | 1_1 1_2 \dots 1_l 0_{l+1} 0_{l+2} \dots 0 \rangle \right ] / \left (
\begin{array}{c} N \\ l \end{array} \right )^{1/2}  
\end{eqnarray}
which are the eigenstates of $\sigma_3$. The difference between excited and unexcited particles is counted by $\sigma _3$, since $_{\rm p} \langle l | \sigma _3 | l \rangle _{\rm p} = l - {1 \over 2} N$. For any $l$ we have \cite{cool}:
\begin{eqnarray}
\sigma ^+ ~ | l \rangle _{\rm p} &=& \sqrt{l+1} \sqrt{N-l}  ~ | l + 1 \rangle _{\rm p} ~ , \nonumber \\ \label{Dicke1}
\sigma ^- ~ | l \rangle _{\rm p} &=& \sqrt{N-(l-1)} \sqrt{l} ~ | l - 1 \rangle _{\rm p} ~ .
\end{eqnarray}
This shows that $\sigma ^{\pm}$ and $\sigma _3$ are represented on $| l
\rangle_{\rm p}$ by the Holstein-Primakoff non-linear boson realization
\cite{Holstein,Shah} $\sigma ^+ = \sqrt{N} S ^+ A_s$, $\sigma ^- = \sqrt{N}
A_s S^- $ with $ \sigma _3 = S^+ S^- - {1 \over 2} N$, $ A_s = \sqrt{1-S^+ S^-
  /N}$, $S^+ | l \rangle _{\rm p} = \sqrt{l+1} | l + 1 \rangle _{\rm p}$, and
$S^- | l \rangle _{\rm p} = \sqrt{l} | l - 1 \rangle _{\rm p}$ for any
$l$. The $\sigma$'s still satisfy the su(2) algebra (\ref{su2}). However, for
$N \gg l$, (\ref{Dicke1}) becomes 
\begin{equation}
\sigma ^{\pm} ~ |l \rangle _{\rm p} = \sqrt{N} S^{\pm} ~ | l \rangle _{\rm p}
~.
\end{equation}
 Consequently, in the large $N$ limit, (\ref{su2}) contracts to the projective algebra e(2) \cite{Shah}
\begin{equation} \label{Dicke3}
\left[ S_3, S^{\pm} \right ] = \pm S^{\pm}  ~, ~~ \left [ S^-, S^+ \right ] = 1 ~,
\end{equation}
in terms of $S^{\pm}$ and $S_3 \equiv \sigma_3$. This means, the $s^\pm$ operators in the cavity case and in the many atom case are formally the same (cf.~(\ref{H})--(\ref{Dicke4})). 

To derive the master equation of a single system, we assume that its state $|\varphi \rangle$ at $t=0$ is known. Moreover, we notice that spontaneously emitted photons leave at a very high speed and cannot be reabsorbed. The free radiation field is hence initially in a state with only a negligible photon population in the optical regime \cite{Hegerfeldt93,Dalibard,Carmichael}. Denoting this state by $|{\cal O} \rangle$, the (unnormalised) state vector of system and bath equals   
\begin{equation} \label{kernel}
|{\cal O} \rangle |\varphi^0_{\rm I} \rangle = |{\cal O} \rangle \langle {\cal O}| ~ U_{\rm I} (\Delta t,0) ~ |{\cal O} \rangle |\varphi \rangle 
\end{equation}
under the condition of {\em no} photon emission in $(0,\Delta t)$. In the interaction picture with respect to $H_0$, this equation can be calculated using second oder perturbation theory, even when $\Delta t \gg 1/\omega$. Doing so, we find 
\begin{equation} \label{kernel2}
|\varphi^0_{\rm I} \rangle = \big[ 1 - A ~ s^+ s^- - B ~ s^- s^+ - C ~ s^{+2} - D ~ s^{-2} \big] ~ |\varphi \rangle
\end{equation}
with
\begin{eqnarray} \label{ABC}
A &=&  \int_0^{\Delta t} \!\! {\rm d}t \int_0^t \!\! {\rm d}t' ~ \sum _{{\bf k}, \lambda} g_{{\bf k} \lambda} \tilde g_{{\bf k} \lambda}^* ~ {\rm e}^{{\rm i}(\omega-\omega_k)(t-t')} ~, \nonumber \\
B &=& \int_0^{\Delta t} \!\! {\rm d}t \int_0^t \!\! {\rm d}t' ~ \sum _{{\bf k}, \lambda} g_{{\bf k} \lambda}^* \tilde g_{{\bf k} \lambda} ~ {\rm e}^{-{\rm i}(\omega+\omega_k)(t-t')} ~, \nonumber \\
C &=& \int_0^{\Delta t} \!\! {\rm d}t \int_0^t \!\! {\rm d}t' ~ \sum _{{\bf k}, \lambda} g_{{\bf k} \lambda} \tilde g_{{\bf k} \lambda} ~ {\rm e}^{{\rm i} (\omega -\omega_k) t + {\rm i} (\omega + \omega_k) t'} ~, \nonumber \\
D &=& \int_0^{\Delta t} \!\! {\rm d}t \int_0^t \!\! {\rm d}t' ~ \sum _{{\bf k}, \lambda} g_{{\bf k} \lambda}^* \tilde g_{{\bf k} \lambda}^* ~ {\rm e}^{- {\rm i} (\omega + \omega_k) t - {\rm i} (\omega - \omega_k) t'}  ~.~~~~~~
\end{eqnarray}
All four parameters could, in principle, be of first order in $\Delta t$ due to the sum over the infinitely many modes of the free radiation field.

In analogy to (\ref{kernel}), the (unnormalised) density matrix of the system in case of an emission equals
\begin{equation} \label{kernel4}
\rho^>_{\rm I} = {\rm Tr}_{\rm R} \left[ \sum _{{\bf k}, \lambda} a ^\dagger _{{\bf k} \lambda} a _{{\bf k} \lambda}  
~ U_{\rm I} (\Delta t,0) ~ \tilde \rho ~ U_{\rm I}^\dagger (\Delta t,0)  \right] 
\end{equation}
with $\tilde \rho = |{\cal O} \rangle \langle {\cal O}| \otimes \rho$ being the initial state of system and bath. Proceeding as above and using again second order perturbation theory, this yields
\begin{equation} \label{kernel5}
\rho^>_{\rm I} = \tilde A ~ s^- \rho s^+ + \tilde B ~ s^+ \rho s^- + \tilde C ~ s^- \rho s^- + \tilde D ~ s^+ \rho s^+ ~. 
\end{equation}
The coefficients $\tilde A$, $\tilde B$, $\tilde C$, and $\tilde D$ are obtained when taking the complex conjugate of the coefficients $A$, $B$, $C$, and $D$ in Eq.~(\ref{ABC}) and extending the integration of the inner integral to $\Delta t$. To obtain relations between these coefficients, we decompose 
\begin{equation}
\int_0^{\Delta t} \!\! {\rm d}t \int_0^{\Delta t} \!\! {\rm d}t' ... =
\int_0^{\Delta t} \!\! {\rm d}t \int_0^t \!\! {\rm d}t' ... + \int_0^{\Delta
  t} \!\! {\rm d}t \int_t^{\Delta t} \!\! {\rm d}t' ... ~.
\end{equation}
Substituting $u=\Delta t - t$ and $u' = \Delta t - t'$ in the second integral (which maps its area onto that of the first one) we find
\begin{eqnarray} \label{CD}
&& \tilde A = 2 {\rm Re} A ~, ~~  \tilde C = C^* + {\rm e}^{- 2 {\rm i} \omega \Delta t} ~ C ~,~~~ \nonumber \\
&& \tilde B = 2 {\rm Re} B ~, ~~ \tilde D = D^* + {\rm e}^{2 {\rm i} \omega \Delta t} ~ D ~.
\end{eqnarray}
Choosing the overall phase of $C$ accordingly \footnote{This is done by adjusting the phases of the states of the free radiation field which affects the $g_{{\bf k} \lambda}$ and $\tilde g_{{\bf k} \lambda}$ in (\ref{H}).}, the parameters $C$, $D$, $\tilde C$, and $\tilde D$ can hence be written as 
\begin{equation}
C = D^* = {1 \over 2}f ~ \gamma_{\rm C} ~, ~~ \tilde C = \tilde D^* = f^* ~
\gamma_{\rm C} ~, 
\end{equation}
with
\begin{equation}
f \equiv  {\rm e}^{{\rm i} \omega \Delta t} ~ \sin(\omega \Delta t) / \omega 
\end{equation}
and with $\gamma_{\rm C}$ being a real but not specified function of $\Delta t$. One can easily check that this notation is consistent with $\tilde D = \tilde C^*$ (cf.~(\ref{ABC})) and with (\ref{CD}). 

Averaging over the subensemble with and the subensemble without photon
emission (cf.~(\ref{kernel2}) and (\ref{kernel5})) at $\Delta t$ hence yields
the density matrix 
\begin{eqnarray} \label{kernel7}
\rho_{\rm I} (\Delta t) &=& \rho - \big[ \big( A ~ s^+ s^- + B ~ s^- s^+) ~ \rho + {\rm h.c.} \big] \nonumber \\
&& \hspace*{-0.4cm} - {1 \over 2} \gamma_{\rm C} ~ \big[ \big( f ~ s^{+2} + {\rm h.c.} \big) ~ \rho + {\rm h.c.} \big] + 2 {\rm Re} A ~ s^- \rho s^+ \nonumber \\
&& \hspace*{-0.4cm} +  2 {\rm Re} B ~  s^+ \rho s^- + \gamma_{\rm C} ~ \big[  f ~ s^+ \rho s^+ + {\rm h.c.} \big] ~.
\end{eqnarray}
In the following we return into the Schr\"odinger picture considering a master equation \footnote{This equation is different from the one in
    G. S. Agarwal, {\em  Quantum Optics}, Springer Tracts of Modern Physics
    Vol. {\bf 70} (Spinger-Verlag, Berlin 1974).}
\begin{eqnarray} \label{deltarho2}
\dot \rho &=& - {{\rm i} \over \hbar} \left[ H_{\rm cond} \rho - \rho H_{\rm cond}^\dagger \right] + {\cal R}(\rho) ~ , \nonumber \\
{\cal R}(\rho) &=& \gamma_{\rm A} ~ s^- \rho s^+ + \gamma_{\rm B} ~ s^+ \rho s^- + \gamma_{\rm C} ~ \big( s^- \rho s^- + {\rm h.c.} \big) ~ , \nonumber \\
H_{\rm cond} &=&- {{\rm i} \over 2} \hbar \big[ \gamma_{\rm A} ~ s^+ s^- + \gamma_{\rm B} ~ s^- s^+ + \gamma_{\rm C} ~ \big( s^{+2} + {\rm h.c.} \big) \big] \nonumber \\
&& +  \hbar {\widetilde \omega} ~ s^+ s^- 
\end{eqnarray}
with $\gamma_{\rm A} = 2 {\rm Re} A/\Delta t$, $\gamma_{\rm B} =  2 {\rm Re} B/\Delta t$, and with ${\widetilde \omega}$ being the shifted bare transition frequency. Checking the validity of (\ref{deltarho2}) can be done easily by returning into the interaction picture and integrating $\dot \rho_{\rm I} (t)$ from zero to $\Delta t$. The result is indeed (\ref{kernel7}).

Before continuing, we remark that the master equation (\ref{deltarho2}) has been derived using second order perturbation theory. This means, it applies to quantum optical systems with the ability to spontaneously emit photons. The assumption of a Markovian bath has been avoided. Instead, we assumed rapidly repeated (absorbing) measurements whether or not a photon has been emitted \cite{Hegerfeldt93,Dalibard,Carmichael}. These measurements constantly reset the free radiation field into $ |{\cal O} \rangle$ and make the system dynamics on the coarse grained time scale $\Delta t$ with $1/\omega \ll \Delta t \ll 1/\gamma$ automatically Markovian. Predictions based on this assumption have already been found in good agreement with actual experiments \cite{Toschek,Schoen}. 

\section{Photon emission from a single quantum system} \label{single}

Let us continue by calculating the probability density $I_\gamma =  {\rm Tr} ({\cal R}(\rho))$ for a photon emission of a system prepared in $\rho$,
\begin{eqnarray} \label{kernel3}
I_\gamma &=& \left \langle \gamma_{\rm A} ~ s^+ s^- + \gamma_{\rm B} ~ s^- s^+ + \gamma_{\rm C} ~ \left( s^{+2} + s^{-2} \right) \right \rangle ~.~~~~
\end{eqnarray}
Using (\ref{bose}) or (\ref{Dicke4}), respectively, and considering the time evolution of the expectation values $\mu_1 \equiv \langle s^+ s^- \rangle$, $\xi_1 \equiv {\rm i} \langle s^{-2} - s^{\dagger +2} \rangle$, and $\xi_2 \equiv \langle s^{-2} + s^{+2} \rangle$, we obtain a closed set of rate equations,
\begin{eqnarray}
&& \dot \mu_1 = - (\gamma_{\rm A} - \gamma_{\rm B}) ~ \mu_1 + \gamma_{\rm B} ~, \nonumber \\
&& \dot \xi_1 = - (\gamma_{\rm A} - \gamma_{\rm B}) ~ \xi_1 + 2 \widetilde \omega ~ \xi_2 ~, \nonumber \\
&& \dot \xi_2 = - (\gamma_{\rm A} - \gamma_{\rm B}) ~ \xi_2 - 2 \tilde \omega ~ \xi_1 - 2 \gamma_{\rm C} ~.
\end{eqnarray}
Setting these derivatives equal to zero, we find that the stationary photon emission rate of a single bosonic system (like an optical cavity or many tightly trapped atoms) is
\begin{eqnarray}
I_\gamma = {2 \gamma_{\rm A} \gamma_{\rm B} \over \gamma_{\rm A} - \gamma_{\rm B}} - {2 \gamma_{\rm C}^2 (\gamma_{\rm A} - \gamma_{\rm B}) \over  4 \widetilde \omega^2 + (\gamma_{\rm A} - \gamma_{\rm B})^2} ~. 
\end{eqnarray}
No photon emissions occur in the absence of external driving only when 
\begin{eqnarray} \label{last}
\gamma_{\rm B} = \gamma_{\rm C} = 0 ~, 
\end{eqnarray}
as it is assumed almost everywhere in the literature \cite{Hegerfeldt93,Dalibard,Carmichael,Werlang}. However, this assumption relies strongly on how the integrals in (\ref{ABC}) are evaluated and whether relations like $\tilde D(\omega) = \tilde C(-\omega)$ are taken into account or not.

\section{Photon emission from a composite quantum system} \label{comp}

Let us now have a look at a large number $N$ of tightly confined atoms inside an optical cavity. The energy of this composite system is the sum of the free energy of both subsystems, their interaction with the free radiation field, and the interaction between the atoms and the cavity field which changes (\ref{H}) into 
\begin{eqnarray} \label{H2}
H_0 &=& \hbar ~ \omega_{\rm c} ~c^\dagger c + \hbar \omega_0 ~ S^+ S^-  + \sum _{{\bf k},\lambda} \hbar \omega _k ~ a ^\dagger _{{\bf k} \lambda} a _{{\bf k} \lambda} ~ , \nonumber \\
H_{\rm int} &=& \sum_{{\bf k}, \lambda} \hbar \big( g_{{\bf k} \lambda} ~ a _{{\bf k} \lambda} + \tilde g_{{\bf k} \lambda} ~ a _{{\bf k} \lambda}^\dagger \big) ~ c^\dagger + \sqrt{N} \hbar ~ \big( q_{{\bf k} \lambda} ~ a _{{\bf k} \lambda} \nonumber \\
&& \hspace*{-0.5cm} + \tilde q_{{\bf k} \lambda} ~ a _{{\bf k} \lambda}^\dagger \big) ~ S^+ + \sqrt{N} \hbar g_{\rm c} ~ \big( c+c^\dagger \big) ~ S^+ + {\rm h.c.} ~~~~
\end{eqnarray}
with $g_{\rm c}$, $g_{{\bf k} \lambda}$, $\tilde g_{{\bf k} \lambda}$,
$q_{{\bf k} \lambda}$ and $\tilde q_{{\bf k} \lambda}$ being coupling
constants. For simplicity, the cavity photon states should be chosen such that
$g_{\rm c}$ becomes real. Proceeding as in the single system case, assuming
the same properties of the bath, and returning into the Schr\"odinger picture
we receive again the master equation (\ref{deltarho2}) but with
\begin{eqnarray} \label{last2}
{\cal R}(\rho) &=& \kappa~ c \rho c^\dagger + N\Gamma ~ S^+ \rho S^- ~, \nonumber \\
H_{\rm cond} &=& \hbar \Big( \widetilde \omega_{\rm c} - {{\rm i} \over 2} \kappa \Big) ~ c^\dagger c + \hbar \Big( \widetilde \omega_0 - {{\rm i} \over 2} N \Gamma \Big) ~ S^+ S^- \nonumber \\
&& + \sqrt{N} \hbar g_{\rm c} ~ \big( c+ c^\dagger \big) \big( S^+ + S^- \big)  ~. 
\end{eqnarray}
Here $\widetilde \omega_{\rm c}$ and $\widetilde \omega_0$ denote the bare atom and cavity frequencies, $\kappa$ is the cavity decay rate, and $\Gamma$ is the decay rate of the excited state of a single atom. The crucial difference to the usual Jaynes-Cummings model \cite{Knight} is the presence of the $c S^-$ and the $c^\dagger S^+$ term in (\ref{H2}) which vanish in the RWA. As we shall see below, these operators result in a non-zero stationary state population in excited states and the continuous emission of photons, even without external driving.

\begin{figure}[t]
\begin{minipage}{\columnwidth}
\begin{center}
{\includegraphics[scale=0.7]{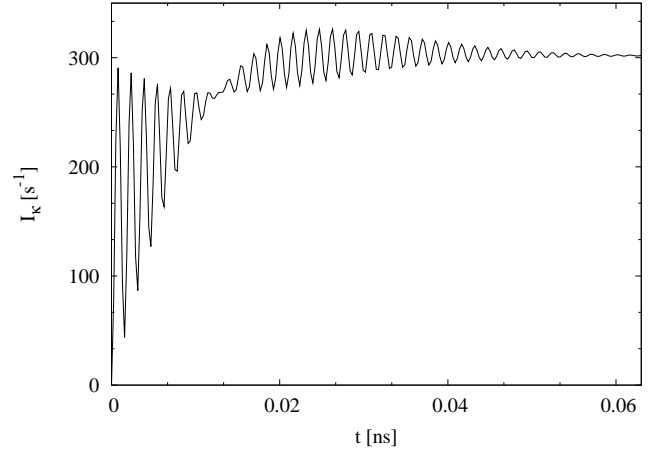}} 
\end{center}
\caption{The cavity photon emission rate $I_\kappa$ as a function of time for $N=10^4$ atoms inside the resonator obtained from a numerical solution of the respective rate equations. Here $\omega_{\rm c} = \omega_0 = 384.2 \cdot 10^{12} \, {\rm s}^{-1}$ ($D_2$ line), $g_{\rm c} = 6.1\cdot 10^8 \, {\rm s}^{-1}$, $\Gamma =  1.9 \cdot 10^7 \, {\rm s}^{-1}$, and $\kappa = 1.3 \cdot 10^{10} \, {\rm s}^{-1}$, as in Ref.~\cite{Trupke}. After being initially empty, the cavity becomes populated --- even in the absence of external driving.} \label{fig2}
\end{minipage}
\end{figure}

To calculate this rate, we take a conservative point of view and neglect $\gamma_{\rm B}$ and $\gamma_{\rm C}$ as suggested in (\ref{last}), since this assures that no emissions occur in the absence of external driving in the single system case. Using (\ref{bose}), (\ref{Dicke4}), (\ref{deltarho2}), and (\ref{last2}) we obtain again a closed set of the rate equations:
\begin{eqnarray}
&&\dot \mu _1 = \sqrt{N} g_{\rm c} \eta _1 - \kappa \mu _1 \, , ~~ 
\dot \mu _2 = \sqrt{N}  g_{\rm c} \eta _2 - N \Gamma \mu _2 ~ , \nonumber \\
&&\dot \eta _1 =  2 \sqrt{N} g_{\rm c} (1 + 2 \mu_2 + \xi_4) + \widetilde \omega_0 \eta _3 + \widetilde \omega_{\rm c} \eta _4 - {\textstyle {1 \over 2}} \zeta \eta_1 ~ , \nonumber \\
&&\dot \eta _2 = 2 \sqrt{N}  g_{\rm c} (1 + 2 \mu_1 + \xi_2) + \widetilde \omega _0 \eta _4 + \widetilde \omega _{\rm c} \eta _3 - {\textstyle {1 \over 2}} \zeta \eta_2 ~ , \nonumber \\
&&\dot \eta _3 = - 2 \sqrt{N}  g_{\rm c} (\xi_1 + \xi_3) - \widetilde \omega _0 \eta _1 - \widetilde \omega _{\rm c} \eta _2 - {\textstyle {1 \over 2}} \zeta \eta _3 ~ , \nonumber \\
&&\dot \eta _4 = - \widetilde \omega _0 \eta _2 - \widetilde \omega_{\rm c} \eta _1 - {\textstyle {1 \over 2}} \zeta \eta_4 ~, \nonumber \\
&&\dot \xi _1 = 2 \sqrt{N} g_{\rm c} \eta_4 + 2 \widetilde \omega_{\rm c} \xi_2 - \kappa \xi_1 ~, \nonumber \\
&&\dot \xi _2 = - 2 \sqrt{N} g_{\rm c} \eta_1 - 2 \widetilde \omega_{\rm c} \xi_1 - \kappa \xi_2 ~, \nonumber \\
&&\dot \xi _3 = 2 \sqrt{N} g_{\rm c} \eta_4 + 2 \widetilde \omega_0 \xi_4 - N \Gamma \xi_3 ~, \nonumber \\
&&\dot \xi _4 = - 2 \sqrt{N} g_{\rm c} \eta_2 - 2 \widetilde \omega_0 \xi_3 - N \Gamma \xi_4
\end{eqnarray}
with $\mu_1 \equiv \langle c^\dagger c \rangle$, $\mu_2 \equiv \langle S^+ S^- \rangle$, $\eta_{1,2} \equiv {\rm i} \langle (S ^- \pm S ^+) (c \mp c^\dagger) \rangle$, $\eta _{3,4} \equiv\langle (S ^- \mp S ^+) (c \mp c^\dagger )\rangle$, $\xi_1 \equiv {\rm i} \langle c^2 - c^{\dagger 2} \rangle$, $\xi_2 \equiv \langle c^2 + c^{\dagger 2} \rangle$, $\xi_3 \equiv {\rm i} \langle S^{-2} - S^{+2} \rangle$, $\xi_4 \equiv \langle S^{-2} + S^{+2} \rangle$, and $\zeta \equiv \kappa + N \Gamma$. Combined with (\ref{kernel3}), the stationary state of these equations yields  the cavity photon emission rate
\begin{eqnarray} \label{IN}
I_{\kappa} &=& {N \zeta \kappa g_{\rm c}^2 \left [ \, 8 \zeta g_{\rm c}^2 + \zeta^2 \Gamma + 4 \Gamma \left ( \widetilde \omega _0 - \widetilde \omega_{\rm c} \right )^2 \, \right] \over 16 \zeta^2 g_{\rm c}^2 \widetilde \omega_0 \widetilde \omega_{\rm c} + 2 \zeta^2 \kappa \Gamma \left ( \widetilde \omega _0 ^2 + \widetilde \omega_{\rm c}^2 \right ) + 4 \kappa \Gamma \left( \widetilde \omega _0^2 - \widetilde \omega_{\rm c}^2 \right)^2} \nonumber \\
\end{eqnarray}
which applies for $N \Gamma, \sqrt{N} g_{\rm c}, \kappa \ll \widetilde \omega_0,\widetilde \omega_{\rm c}$. For example, the parameters of the recent cavity experiment with $^{85}$Rb $[4]$ combined with $N=10^4$ are expected to result in an $I_\kappa$ as large as $I_\kappa =301~\rm{s}^{-1}$ which can be detected experimentally (cf.~Fig.~\ref{fig2}).

\section{Conclusions} \label{conc}

Let us now discuss the same topic in a more physical way. Among the reasons usually produced to justify the RWA, there is the one of the time scale, since the counter-rotating terms oscillate very rapidly so that their contribution to the dynamics remains in general negligible with respect to the resonating ones \cite{Grynberg}. Other authors apply the RWA in order to preserve quantum numbers and energy \cite{Milonni,Schleich}, since this approximation drops all the processes transferring energy between non-resonating modes. However, in Nature, there exist processes, where there is a redistribution of energy among different system degrees of freedom making possible some amounts of system self-organization. In particular, one could examine the possibility of concentrating the total energy of the system into a subset of degrees of freedom producing a decrease of its entropy, which in order to avoid a violation of the second law of thermodynamics, would compel the release of energy to the environment, thus keeping the free energy constant. This is possible, of course, only when the system is open.

In this paper, we consequently examined the situation occurring when the 
counter-rotating terms are not dropped. The avoidance of the RWA and the 
consequent more exact solution of the dynamics of the system under consideration gives rise to non-trivial consequences. As a matter of fact, the mechanism of concentration of energy on a subset of degrees of freedom could help the understanding of the hitherto rather mysterious processes of self-organization. The mathematical analysis we have done above shows indeed that in a quantum system a leakage of energy can occur among 
different degrees of freedom. This leakage is not necessarily triggered by an external pump of energy, but
could be also triggered by the virtual photons coming from the quantum 
vacuum as, e.g., it occurs in the Casimir effect or in the Lamb shift. From the 
standpoint of the receiving system, the origin of the triggering energy is not important as far as 
the balance between the variations of energy and entropy is satisfied so 
to keep the free energy constant. In this respect, we recall that the ratio between these variations is just 
the temperature, as required by the thermodynamic definition: 
\begin{eqnarray}
k_{\rm B} T &=& {{\mathrm d}U \over {\mathrm d}S} ~.
\end{eqnarray}

The interplay between the microscopic quantum dynamics and the thermal properties certainly deserves further analysis, which, however, is out of the scope of the present paper. Here we are limiting ourselves to a specific physical picture, which however, does not exclude other physical scenarios, as, for example, the conversion of 
energy from the thermal bath phonons to leaking photons (as it occurs in the laser 
cooling mechanism). More
appealing could be the occurrence of a dynamics, where a system is able 
to reach a state having
a lower energy jumping over a separating barrier with the help of 
virtual photons coming from the vacuum. The possibility of realization 
of these scenarios needs of course further studies. First indications 
along the above lines can be found in the literature \cite{Kurcz_NJP,hegerfeldt2,Zubairy,Werlang}. In all these examples, the system dynamics is irreversible.

In summary, we derived the master equation for a single bosonic system (an optical cavity and a large number of tightly confined particles) without making any approximation other than the usual dipole and Born approximation. We find that the effect of the counter-rotating terms in the interaction between a quantum optical system and its free radiation field might be annihilated by environment-induced measurements whether or not a photon has been emitted. Assuming that this is the case, we then show that these measurements cannot suppress the interaction between a large number of atoms and an optical cavity. The result is the continuous leakage of photons through the resonator mirrors, even in the absence of external driving. For sufficiently many atoms, a relatively strong signal might be created. Its dependence on the system parameters can be verified experimentally using optical cavities like those described in \cite{Trupke,Reichel,Esslinger}. We recognize that in order to better understand the physical mechanisms responsible for the mathematical results presented in this paper, some more work is needed which is in our plans for future publications.

\vspace*{0.05cm}

\noindent {\em Acknowledgement.} A. B. acknowledges a James Ellis University Research Fellowship from the Royal Society and the GCHQ. This work was supported in part by the UK Research Council EPSRC, the EU Research and Training Network EMALI, University of Salerno, and INFN.


\begin{thebibliography}{99}
\bibitem{Kurcz_NJP}
A. Kurcz, A. Capolupo, A. Beige, E. Del Giudice, and G. Vitiello, {\em Rotating wave approximation and entropy}, New J. Phys. (submitted); arXiv:1001.3944.

\bibitem{Bartana}
A. Bartana, R. Kosloff, and D. J. Tannor, J. Chem. Phys. {\bf 106}, 1435 (1997).

\bibitem{Hegerfeldt93}
G. C. Hegerfeldt, Phys. Rev. A {\bf 47}, 449 (1993).

\bibitem{Dalibard}
J. Dalibard, Y. Castin, and K. M{\o}lmer, Phys. Rev. Lett. {\bf 68}, 580 (1992).

\bibitem{Carmichael}
H. Carmichael, {\em  An Open Systems Approach to Quantum Optics}, Lecture Notes in Physics, Vol. {\bf 18} (Springer, Berlin, 1993).

\bibitem{Wineland}
D. J. Wineland and W. M. Itano, Phys. Rev. A {\bf 20}, 1521 (1979).

\bibitem{agarwal}
G. S. Agarwal, Phys. Rev. A {\bf 4}, 1778 (1971).

\bibitem{knight}
P. L. Knight and L. Allen, Phys. Rev. A {\bf 7}, 368 (1973).

\bibitem{hegerfeldt2}
G. C. Hegerfeldt, Phys. Rev. Lett. {\bf 72}, 596 (1994).

\bibitem{Zubairy}
H. Zheng, S. Y. Zhu, and M. S. Zubairy, Phys. Rev. Lett. {\bf 101}, 200404 (2008).

\bibitem{Werlang}
T. Werlang, A. V. Dodonov, E. I. Duzzioni, and C. J. Villas-Boas, Phys. Rev. A {\bf 78}, 053805 (2008).

\bibitem{Trupke}
M. Trupke, J. Goldwin, B. Darquie, G. Dutier, S. Eriksson, J. Ashmore, and
E. A. Hinds, Phys. Rev. Lett. {\bf 99}, 063601 (2007).

\bibitem{Reichel}
Y. Colombe, T. Steinmetz, G. Dubois, F. Linke, D. Hunger, and J. Reichel, Nature {\bf 450}, 272 (2007).

\bibitem{Esslinger}
F. Brennecke, S. Ritter, T. Donner, and T. Esslinger, Science {\bf 322}, 235 (2008).

\bibitem{Holstein}
T. Holstein and H. Primakoff, Phys. Rev. {\bf 58}, 1098 (1940).

\bibitem{Shah}
M. N. Shah, H. Umezawa, and G. Vitiello, Phys. Rev. B {\bf 10}, 4724 (1974).

\bibitem{SL}
A. Kurcz, A. Capolupo, and A. Beige, New J. Phys. {\bf 11}, 053001 (2009).

\bibitem{review}
D. Leibfried, R. Blatt, C. Monroe, and D. Wineland, Rev. Mod. Phys. {\bf
  75}, 281 (2003).

\bibitem{cool}
A. Beige, P. L. Knight, and  G. Vitiello, New J. Phys. {\bf 7}, 96  (2005).

\bibitem{Toschek}
A.~Beige~and~G.~C.~Hegerfeldt,~Phys.~Rev.~A~{\bf 53},~53~(1996).

\bibitem{Schoen}
C. Sch\"on and A. Beige, Phys. Rev. A {\bf 64}, 023806 (2001).

\bibitem{Knight}
B. W. Shore and P. L. Knight, J. Mod. Opt. {\bf 40}, 1195 (1993).

\bibitem{Grynberg}
C. Cohen-Tannoudji, J. Dupont-Roc, and G. Grynberg, {\it Atom-Photon
  Interactions} (Wiley-VCH Verlag GmbH \& Co. KGaA, Weinheim, 2004).

\bibitem{Milonni}
P. W. Milonni, {\it The quantum vacuum} (Academic Press Limited, London, 1994).

\bibitem{Schleich}
W. P. Schleich, {\it Quantum Optics in Phase Space} (Wiley-VCH Verlag GmbH,
Berlin, 1994).
\end{thebibliography}
\end{document}